\documentclass[]{spie}  %>>> use for US letter paper
\usepackage{amsmath,amsfonts,amssymb}
\usepackage{deluxetable}
\usepackage{graphicx}
\usepackage[colorlinks=true, allcolors=blue]{hyperref}

\title{Calibration of MEMS DM actuator gains using a Zernike wavefront sensor on the HiCAT testbed and implications for Habitable Worlds Observatory operations}

\author[a]{Sarah Steiger}
\author[b]{Raphaël Pourcelot}
\author[a]{Rémi Soummer}
\author[ ]{Emiel H. Por}
\author[a]{Laurent Pueyo}
\author[c]{Iva Laginja}
\author[a]{Nathan Scott}

\affil[a]{Space Telescope Science Institute, Baltimore MD, USA}
\affil[b]{Max Planck Institute for Astronomy, Heidelberg, Germany}
\affil[c]{Université Côte d'Azur, Observatoire de la Côte d'Azur, CNRS, Laboratoire Lagrange, Nice, France}

\authorinfo{Further author information: (Send correspondence to S.S.)\\S.S.: E-mail:ssteiger@stsci.edu}

\begin{document} 
\maketitle

\begin{abstract}
Deformable mirrors (DMs) are a key component of coronagraph instruments performing adaptive optics on the ground, and for future space observatories such as HWO and Roman CGI. Here, they will be used as part of the wavefront sensing and control system to ``dig a dark zone'' – remove residual stellar light to create a high-contrast region in the focal plane where faint companions can be detected and characterized. To reach the deep contrasts needed to directly image cool or reflected light planets ($<$1e-8) accurate calibration of the DM actuator gain is essential as picometer differences between the expected and realized DM surface can significantly degrade dark zone (DZ) digging efficiency. This increases the overheads needed to achieve a DZ and critically places more stringent requirements on observatory stability as the level of tolerable drift decreases with the required iterations to achieve the desired contrast. Furthermore, DM gain varies with actuator stroke, necessitating rapid, \textit{in situ} gain map recalculations to maintain DZ digging efficiency over time. Zernike wavefront sensors (ZWFS) are well-suited for this task as they efficiently provide picometer-level sensitivity and will likely already be included on board as part of a low order wavefront sensor for HWO. Using the ZWFS for DM calibrations however will have significant operational implications for the observing strategy of HWO and so must be investigated early. Here we present results from the HiCAT testbed at STScI where we calibrated gain maps for our Boston Micromachines 952-actuator micro electromechanical (MEMS) DMs using both a Fizeau interferometer and a ZWFS to compare performance. With the ZWFS we compute a gain map using both local linear fits around a given DM solution, and present a formalism for deriving more complex quadratic solutions which are more computationally intensive, but accurate over most of the dynamic range of each actuator. We then use these techniques to calibrate the DMs on the HiCAT testbed and  show increased DZ digging efficiency with the new gain map and better contrast performance moving from 14 to 16 bit control electronics as enabled by these calibrations.

\end{abstract}

\keywords{deformable mirrors, Zernike wavefront sensor, testbed, Habitable Worlds Observatory}

\section{INTRODUCTION}
\label{sec:intro} 

The \emph{Pathways to Discovery in Astronomy and Astrophysics for the 2020s} decadal survey (``Astro2020'')\cite{2021astro2020} recommends the development of a large UVOIR space telescope, now known as the Habitable Worlds Observatory (HWO)\cite{2024HWOFeinberg}. One of the main science drivers of HWO will be exploring the Habitable Zones of $\sim$ 100 nearby star systems and taking spectra of $\sim 25$ rocky planets to search for biosignatures that could be indicative of life. This will require the telescope and coronagraph system to disentangle planetary signals that are tens of billions of times fainter than the stars they orbit placing tight constraints on the performance of all observatory systems. A key component of the coronagraph on HWO will be deformable mirrors (DMs) which are active optics onto which arbitrary shapes can be applied using electronically controlled actuators behind the mirror surface. On HWO these will be used for wavefront sensing and control by calculating and applying the surface shape which will best destructively interfere coherent starlight in the regions where incoherent planet light may be hiding in a process often referred to as ``digging a dark zone (DZ)''. This technique will be proven for the first time in space with the Roman Coronagraph Instrument (CGI)\cite{2020SPIECGI} and will be imperative if HWO is to reach the contrasts required to achieve its main science objectives. 

To reach the deep contrasts needed to directly image cool reflected light planets with HWO or Roman CGI ($<$1e-8), accurate calibration of the DMs, including flat maps and actuator gains, is essential as picometer differences between the expected and realized DM surface can significantly degrade DZ digging efficiency. This increases the overheads needed to achieve a DZ and critically places more stringent requirements on observatory stability as the level of tolerable drift decreases with the required iterations to achieve the desired contrast. Furthermore, DM gain varies with actuator stroke, which could necessitate rapid, \textit{in situ} gain map recalculations to maintain DZ digging efficiency over time. Zernike wavefront sensors (ZWFS) are well-suited for this task as they provide a relatively simple and photon-efficient solution to reach picometer-level sensitivity and will likely already be included on board as part of a low order wavefront sensor for HWO. Using the ZWFS for DM calibrations will therefore require no additional specialized optics, but could have significant operational implications for the observing strategy of HWO and so must be investigated early. 

In this paper we first present the mathematical formalism for calculating DM actuator gains with a ZWFS in Section \ref{sec:gain_math}. Here we explore both a linear model, and a quadratic model which together well describes the measured gain responses of some of the most common DM technologies. We then show results in Section \ref{sec:hicat_map_calc} from the  High-Contrast Imager for Complex Aperture Telescopes (HiCAT)\cite{2024hicat11} testbed at STScI where we calibrated both DM flat maps and linear gain maps for one of our Boston Micromachines 952-actuator micro-electromechanical (MEMS) ``kilo-DMs'' using a ZWFS and compare those to our old DM flat and gain maps that were calculated using a Fizeau interferometer. This work was done as part of a DM electronics upgrade for HiCAT moving from 14-bit to 16-bit control electronics and so gain maps and flat maps at both of these bit depths were generated. In Section \ref{sec:hicat_dzs} we use these new DM flats and gain maps to dig dark zones on HiCAT showing increased DZ digging efficiency with the new gain maps and better contrast performance moving from 14 to 16-bit control electronics as enabled by these calibrations.

\section{GAIN MAP CALCULATIONS}
\label{sec:gain_math}

Before we can perform our calibrations, we first need a mathematical model to describe the expected DM response as a function of input actuator voltage. A common DM technology are MEMs DMs, of which the most prevalent commercial options are from Boston Micromachines (BMC) which will be the subject of this work in Sections \ref{sec:hicat_map_calc} and \ref{sec:hicat_dzs}. The response of these DMs exhibit a well known quadratic relationship with input voltage\cite{2006OExprEvans, 2018magaoxDMchar} however 
% which can be approximated as linear in a small range around a given reference voltage to simplify computations. 
this behavior is not ubiquitous - another popular DM manufacturer ALPAO uses magnetic voice-coil actuators and instead exhibits a linear response with voltage\cite{2018magaoxDMchar}.

To accommodate these different DM technologies, we present two mathematical solutions for calculating the actuator gains given a set of input DM commands and optical path differences (OPDs) as measured by a Zernike WFS. The first is a linear solution which is easily invertible, but may only be valid in a small range around a given reference voltage if the DM response is quadratic such as is the case for the BMC DMs. The second quadratic solution would well describe these DMs over their whole dynamic range, but is not easily invertible and may also not be necessary should future HWO operations allow for \textit{in situ} gain recalibration as will be discussed in Section \ref{sec:conclusions}. 

\subsection{Linear Gain Maps}
\label{sec:linear_gain_math}
We will take \textbf{V} to be a list of vectors of random DM actuator positions (in volts) of dimension $m \times N_{act}$. Here $m$ is an arbitrary number of iterations and $N_{act}$ are the number of DM actuators. In practice, $m$ just needs to be selected large enough so as to fully sample all of the actuators and inter-actuator interactions. 

We will then apply each series of $m$ random actuator positions and measure the optical path difference (OPD) for each actuator, for each iteration. This is implemented by taking the double difference image for each set of positions to get the ZWFS response and then using the \texttt{pyzelda}\cite{2018pyzelda} package to convert these images into OPDs. From each OPD map, We extract a piston value for each actuator, which lets us build the vector list \textbf{O} which has the same dimensions as \textbf{V}. For this linear approximation, we can then write

\begin{equation}
    \textbf{O} = \textbf{G}\textbf{V}
\end{equation}

where \textbf{G} is the linear gain map of dimension $N_{act} \times N_{act}$. We can then calculate this linear gain as a straightforward matrix multiplication

\begin{equation}
    \textbf{G} = \textbf{V}^{-1}\textbf{O}
\end{equation}

Where $\textbf{V}^{-1}$ is the pseudo-inverse of \textbf{V} and \textbf{G} is the $N_{act} \times N_{act}$ linear gain map including all of the inter-actuator interaction terms, such as the actuator response functions. 

\subsection{Quadratic Gain Maps}
If the DM response follows a quadratic relationship with voltage, a linear approximation implicitly assumes that each gain is calculated around a given reference voltage ($V_{ref}$). As a DZ solution is found, this voltage for each actuator will change and the relevant gain will therefore also change. We can calculate a gain solution independent of reference voltage by defining an additional quadratic gain term \textbf{M} and solving the following for both \textbf{M} and the linear gain term \textbf{G}, as defined in the previous section.  

\begin{equation}
    \textbf{O} = \textbf{M}\left(V-V_{ref}\right)^2 + \textbf{G}\left(V-V_{ref}\right)
\label{eq:quad_gain}
\end{equation}

Here \textbf{O} are the calculated $m \times N_{act}$ OPDs, \textbf{M} is an $N_{act} \times N_{act}$ quadratic gain matrix, \textbf{G} is the $N_{act} \times N_{act}$ linear gain matrix, and \textit{V} are the $m \times N_{act}$ commanded DM positions (in volts). This is not easily invertible and so calculating an analytic solution is left for future work. We will proceed by calculating \textbf{M} and \textbf{G} numerically.

For simplicity we will let $\textbf{V} = V - V_{ref}$ and define the matrix \textbf{X} as the concatenation of \textbf{V} and \textbf{V}$^2$

\begin{equation}
    \textbf{X} = \textbf{V}^2 | \textbf{V}
\end{equation}

we can then use a linear least-squared fitting routine to solve the following for $\left(\textbf{M}|\textbf{G}\right)$

\begin{equation}
    \textbf{X}\left(\textbf{M}|\textbf{G}\right) = \textbf{O}
\end{equation}

These two terms will describe the actuator gain at any reference voltage and so is accurate over the whole dynamic range of the quadratic DM. While this can be powerful, in many instances the DM will be operated very close to $V_{ref}$ and so the linear gain approximation for which we have an analytic solution is sufficient. This is especially true if \textbf{G} can be recalculated during operations as will be explored in Section \ref{sec:conclusions}. 

If we assume that the inter-actuator coupling is negligible, we can choose to only use the diagonal terms. In this case Eq. \ref{eq:quad_gain} is invertible and an analytical solution can be found. 

For a given actuator $i$, Equation \ref{eq:quad_gain} becomes

\begin{equation}
    O_i = G_i (V_i - V_{ref, i})^2 + M_i (V_i - V_{ref, i})
\end{equation}

and finally

\begin{equation}
    V_i - V_{ref, i} = \frac{-M_i + \sqrt{M_i^2-4G_iO_i}}{2G_i}
\end{equation}

This ignores the inter-actuator interactions but is valid over all voltages and so which is best to use will depend on the specific WFS\&C scenario. 

\begin{figure}
    \centering
    \includegraphics[width=0.6\linewidth]{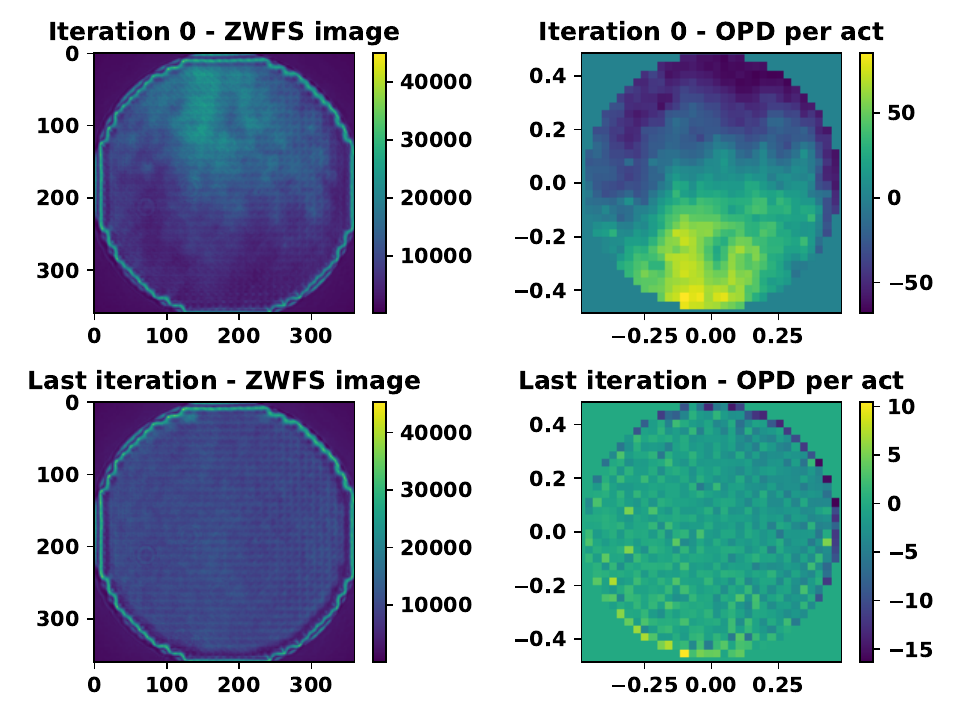}
    \caption{Top left: Starting image on the ZWFS camera. Top right: Calculated OPD of the first closed-loop flat iteration. Bottom left: Final image on the ZWFS camera with the flat solution applied. Bottom right: Final OPD for the last two closed-loop iterations. Note that the OPDs are flipped vertically as compared to the ZWFS images.}
    \label{fig:flatmap}
\end{figure}

\section{Calibrating a BMC DM on HiCAT with a Zernike WFS}
\label{sec:hicat_map_calc}

On HiCAT, we have a ZWFS as part of our low order wavefront sensing (LOWFS) arm which uses rejected light from the focal plane mask for sensing. Typically HiCAT deploys 2 BMC 952 actuator kilo-DMs to control both phase and amplitude over a circular DZ region. To prevent redundancies in the calculation of our flat maps that can arise with having two DMs in the optical chain, we removed our out-of-pupil-plane DM for these calibrations. Once the in-pupil DM is calibrated and a true flat state determined, the other DM will then be returned and calibrated using these same procedures. HiCAT also normally uses an IrisAO DM to create a segmented aperture which was replaced by a flat mirror for these calibrations resulting in a monolithic aperture.

\subsection{Zernike Flat Maps}

To prepare for gain calibrations, we first use the ZWFS to calculate closed-loop DM flat maps, see Figure \ref{fig:flatmap}. To calculate these, we must first measure the OPD of the DM actuators with the ZWFS. This involves taking a dark and a clear pupil image (an image on the ZWFS where the PSF is positioned off of the Zernike dimple) to use as calibration images. From the ZWFS images, the OPDs can then be calculated using the \texttt{pyzelda Sensor.analyze()} utility taking the ZWFS images, darks, and clear pupil images as input. Once we have these OPDs, we can then run phase conjugation by iteratively applying the opposite DM shape (multiplied by a controller gain parameter) for a user specified number of iterations to flatten the DM. This method gives the flattest shape on the ZWFS camera, but could cause the flat to include some non common path aberrations (NCPA) that are present between the DM and the ZWFS in the LOWFS arm, as well as alignment errors of the mask that would appear as tip, tilt and defocus. If this NCPA OPD is known, then it could be subtracted from the measured OPD before applying the control resulting in a ``true'' DM flat.      

\begin{figure}
    \centering
    \includegraphics[width=0.4\linewidth]{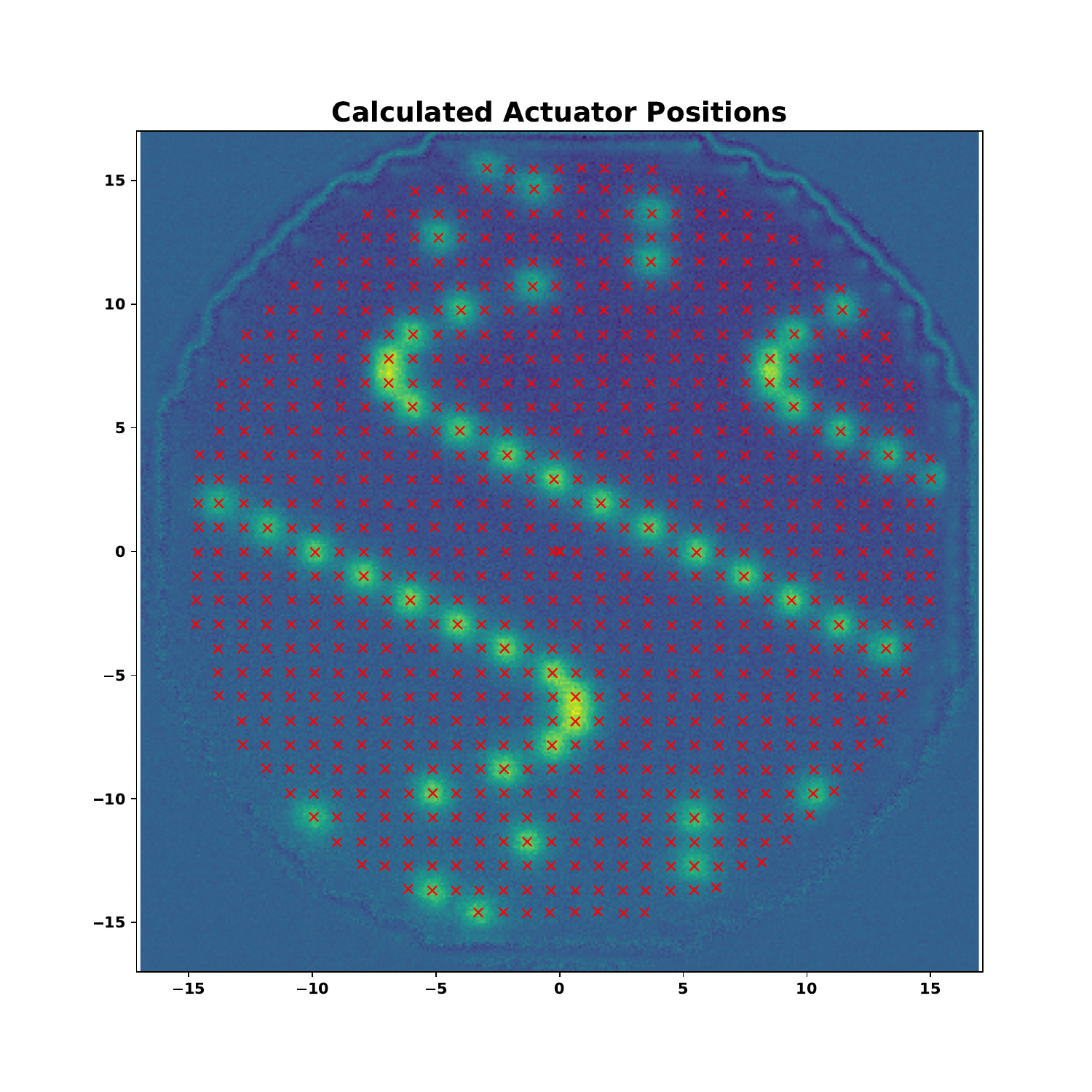}
    \caption{ZWFS image of the BMC DM where actuator positions are identified by performing single actuator pokes and are labeled with red x's. This is a composite image of many individual pokes which is why there are many bright regions (pushed actuators) in the image.}
    \label{fig:actuator_positions}
\end{figure}

Flattening the DMs with the ZWFS also helps us to generate a mapping of the actuator positions on the ZWFS camera which will be important for generating the OPD matrix for the gain maps later since they are evaluated per actuator. An example of this can be found in Figure \ref{fig:actuator_positions} which is a composite of many DM single actuator pokes that show up as bright spots on the image, as well as the red x's that denote the identified actuator positions.

\subsection{Linear Gain Maps}
\label{sec:linear_gain_calc}

For this work we choose to use the linear gain map solution as described in Section \ref{sec:linear_gain_math} due to us having a full analytic solution, and the fact that we can recalculate the gain map \textit{in situ} should the actuators move out of the initial linear approximation during DZ digging. This is enabled by the fact that we are using the light in the LOWFS arm of our testbed and so these calibrations do not require changing any physical hardware in the testbed to complete.  

Following the procedure in Section \ref{sec:linear_gain_math}, we used a series of $m=10,000$ random actuator positions with amplitudes sampled from a Gaussian distribution with a mean of 2 nm and standard deviation of 1 to generate the matrix \textbf{V}. For our implementation, we are setting our DM gains to 1 to collect the data needed to generate the new gain maps and so \textbf{V} is also in units of nm as well as volts. As with the flat maps, the OPD matrix \textbf{O} is calculated by first taking a dark and a clear pupil image to use as calibration images for the ZWFS. Then a positive and negative of each $N_{act}$ random actuator position is applied to generate a positive and negative measure of the OPD using \texttt{pyzelda}. The difference of these two OPDs is then taken and divided by 2. The actuator values that we extract are computed by taking the average of the OPD over a circular region of interest, calculated per actuator and centered on the positions identified during calibration, denoted by the red x's on Figure \ref{fig:actuator_positions}. This results in a single OPD value for each actuator which is then appended to the matrix \textbf{O}. We can then calculate the pseudo-inverse of \textbf{V} using Tikhonov regularization and multiply it by \textbf{O} to finally get the linear gain matrix \textbf{G}, which can be seen in Figure \ref{fig:gain_matrix}. The diagonal of this gain matrix represents the single value gains for each actuator not taking into account any influence functions or other inter-actuator interactions. These more complex effects are encapsulated in the off-diagonal terms (see Figure \ref{fig:gain_matrix}, first two panels).

\begin{deluxetable}{ccc}
\tablewidth{0pt}
\tabletypesize{\scriptsize}
\tablecaption{HiCAT testbed and DZ parameters \label{table:hicat_params}}
\tablehead{
\colhead{Parameter}  & \colhead{Value} & \colhead{Description}
}
\startdata
Aperture  & 19.57 mm (circular) & Aperture size\\
$\lambda$  & 637 nm & Wavelength\\
$\Delta \lambda$  & 20 nm & Bandpass \\
IWA  & 5.4 $\lambda/D_{lyot}$ & DZ inner working angle\\
OWA  & 9.2 $\lambda/D_{lyot}$ & DZ outer working angle \\
FPM  & knife-edge & Focal plane mask type\\
Lyot Stop  & 15 mm & Lyot stop size\\
\hline 
\enddata
\end{deluxetable}

\section{Dark Zone Digging Efficiency and Contrast Results}
\label{sec:hicat_dzs}

We will now quantify the impact these new gain maps have on digging a DZ with the HiCAT testbed. In addition to the monolithic aperture and single in-pupil DM, we implemented several modifications to the HiCAT testbed relative to the configuration described in Soummer et al. (2024)\cite{2024hicat11}. These temporary changes were made to facilitate improved calibration of individual components, such as those described in this work. For the experiments presented here, the testbed was operated in monochromatic light at 637 nm using a 70 mW Thorlabs S4FC Fabry–Perot laser source. We used a PAPLC coronagraph\cite{2020PorPAPLC} with a knife-edge FPM and a 15 mm Lyot stop. A summary of the current HiCAT hardware status can be found in Table \ref{table:hicat_params}. For wavefront sensing and control, we are using electric field conjugation (EFC) with pairwise probing\cite{2007Giveon} using single actuator probe pairs. 

HiCAT uses the Control and Automation for Testbeds Kit 2 (CATKit2) \cite{emiel_h_por_2026_20843325, 2026catkit2hci} software infrastructure to communicate to its hardware and all HiCAT experiments are run through Python scripts for consistency and repeatability. For all of the following experiments, this allows us to only change either the gain map file used for the DM and/or whether the 14 or 16-bit control electronics were used. All other experiment parameters such as control loop gain, regularization schedule, probe number and amplitude, and exposure time were kept exactly the same. This was done to isolate the impact of the gain maps and control bit depth, but may not result in the most optimal scientific performance for all scenarios.   

\subsection{Old gain map vs. new diagonal-only gain map}
\begin{figure}
    \centering
    \includegraphics[width=1.0\linewidth]{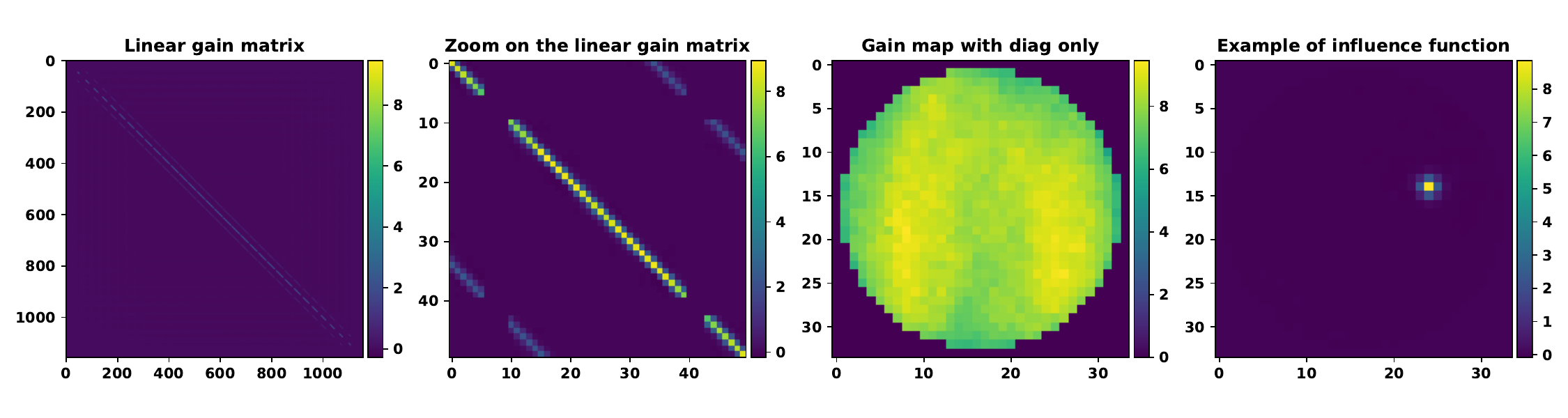}
    \caption{From left to right: Total calculated linear gain matrix \textbf{G}, zoom-in on the diagonal of the linear gain matrix, total gain map using only the diagonal values of the linear gain matrix reshaped into a DM map, an example of a single actuator influence function extracted as a line of \textbf{G} and reshaped into a gain map.}
    \label{fig:gain_matrix}
\end{figure}

\begin{figure}
    \centering
    \includegraphics[width=0.8\linewidth]{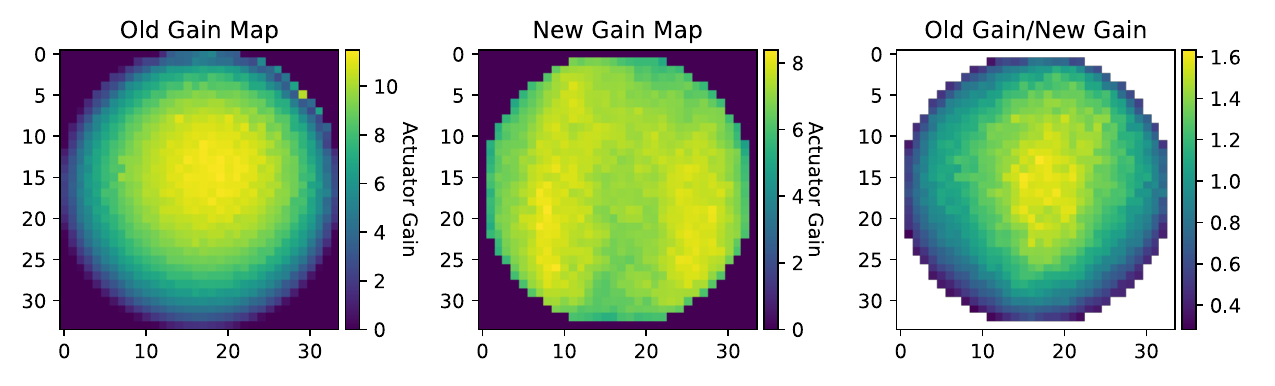}
    \caption{Comparison of the old gain map calculated using the Fizeau interferometer (left), and the diagonal terms of the new linear gain map calculated using the ZWFS (middle). The panel on the right shows the ratio of these two maps where the calculated gains can be seen to differ by up to a factor of 2.}
    \label{fig:gain_map_compare}
\end{figure}

Previously, gain maps for the HiCAT DMs were generated by removing the DMs from the testbed and placing them in front of an Accufiz 4D Fizeau interferometer. Zernike polynomials were then applied to the DM and the actuator positions were measured with the interferometer to generate the gain map. To quantify the improvements of using our new ZWFS method, we first we replaced our old gain map with the gain map generated in Section \ref{sec:linear_gain_calc}. In this instance we are using only the diagonal of the gain matrix (Figure \ref{fig:gain_matrix}, third panel). This was done to be compatible with our existing software infrastructure which can accommodate a 2D actuator gain map, but does not yet contain any information about the influence functions or other inter-actuator correlations which will be explored in future work. A direct qualitative comparison of the new calculated gain map and the old gain map can be found in Figure \ref{fig:gain_map_compare}.  

We then performed identical runs of EFC with pairwise probing using the old and new gain maps to directly compare DZ digging efficiency which can be found in Figure \ref{fig:combined_dz_runs}, left panel. Here we see significant improvements in the convergence of our EFC loop with the new gain map resulting in less iterations required to reach the final achieved contrast. The plateaus that can be seen in the contrast vs. iteration plot are due to the scheduled changing of the regularization parameter for the Jacobian. This indicates that with dedicated optimization, the new gain map could converge even faster.    

% While this work showed significant improvements in our EFC performance with better DM gain performance this was also not taking into account any inter-actuator correlations. An implementation of the full gain matrix (Figure \ref{fig:gain_matrix}, first panel) including these calculated correlations and influence functions will be left for future work.   

\subsection{14 vs. 16-bit DM control electronics}

\begin{figure}
    \centering
    \includegraphics[width=0.95\linewidth]{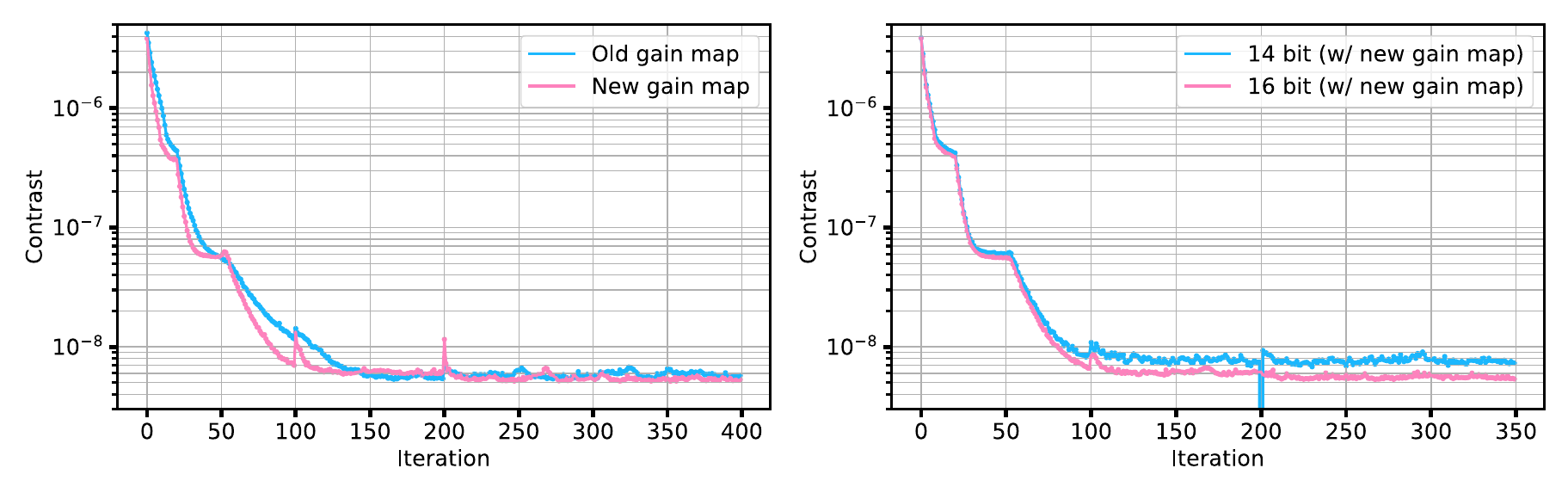}
    \caption{Left: Contrast vs. EFC iteration plot comparing the old gain map (blue) with the new diagonal-only linear gain map (pink). With the new gain map, the convergence is faster due to better control agency. Right: Contrast vs. EFC iteration plot comparing the 14-bit DM electronics (blue) with the 16-bit electronics (pink) where in both cases the new linear gain maps were used. In this case we see that we were limited by DM quantization at 14-bits leading to a higher contrast floor as compared to the 16-bit case. The dip at iteration 200 in the 14-bit plot is an artifact caused by the scheduled acquisition of an unocculted PSF image for contrast calibration. For all of these cases, all other EFC parameters were kept identical.}
    \label{fig:combined_dz_runs}
\end{figure}

The ability to non-invasively calculate gain maps on HiCAT facilitates our ability to make DM electronics upgrades without the overheads or risks of needing to handle the DMs to move them out of the optical path. This was tested when we upgraded our DM control electronics from 14-bits to 16-bits by swapping out the PCIe card as provided by Boston Micromachines. We elected to make this upgrade as simulations indicated that the DM quantization was limiting the final contrast in our DZ at $\sim$1e-8 for our typical testbed configuration. We calculated new gain maps for both of these bit-depths and did a direct comparison running the same EFC loop for both cases, the results of which can be found in Figure \ref{fig:combined_dz_runs}, right panel. Averaging six EFC runs for each bit depth, we see an improvement in the final DZ contrast of 22\% with the 14-bit control electronics reaching a final contrast of $6.8\pm$0.6 $\times 10^{-9}$  and the 16-bit control electronics reaching a final contrast of $5.3\pm 0.3 \times 10^{-9}$.   

\section{Conclusions}
\label{sec:conclusions}

We have  presented the mathematical formalism and procedure to calculate DM flat and gain maps using a ZWFS as part of a coronagraph's LOWFS system. We also demonstrated how these gain and flat maps result in faster EFC convergence on the HiCAT testbed and how they enabled a DM electronics upgrade from 14 to 16-bits. This change improved our realized contrast floor by 22\% from $6.8\pm$0.6 $\times 10^{-9}$ to $5.3\pm 0.3 \times 10^{-9}$ for our PAPLC coronagraph when changing no other testbed or EFC parameters.

Future work will include utilizing the full gain matrix, including all inter-actuator correlations, to assess how this further impacts EFC loop convergence in comparison to the diagonal-only linear gain maps. We are also interested in demonstrating ``on the fly'' gain map recalculations where a partial dark zone is achieved and then a new gain map solution is calculated around those given DZ voltages using the ZWFS. The goal would be to further improve DZ digging efficiency for the final few iterations which can be the most time intensive, especially in low photon regimes. This is an important consideration for HWO as the low flux rates in the DZ as the contrast approaches 1e-10 will mean that very long exposure times are needed to properly sense the electric field for these final iterations. Decreasing the total number of iterations required to reach the final desired contrast could thus significantly decrease the total overheads for DZ digging and increase the time available for science observations. 

DZ digging will be a key aspect of HWO operations and scheduling, but will also be a source of significant overheads for exo-Earth imaging and characterization. By having a way to better calibrate the DMs in-flight (and during DZ digging itself) we hope to provide a path for reducing these overheads and have a more scientifically productive mission overall.

\acknowledgments      
 
The HiCAT testbed has been developed over the past 10 years and benefited from the work of an extended collaboration of over 50 people. This work also benefited from the CATKit2 high-contrast imaging collaboration which originated at the HiCAT testbed at STScI \cite{2024hicat11, 2026catkit2hci}. This work was supported in part by the National Aeronautics and Space Administration under Grant 80NSSC19K0120 issued through the Strategic Astrophysics Technology/Technology Demonstration for Exo-planet Missions Program (SAT-TDEM; PI: R. Soummer), and under Grant 80NSSC22K0372 issued through the Astrophysics Research and Analysis Program (APRA; PI: L. Pueyo). S.S. acknowledges support from an STScI Postdoctoral Fellowship. R.P. was partly supported by Grant 
80NSSC2\-2K0372 issued through the Astrophysics Research and Analysis Program (APRA; PI: L. Pueyo). E.H.P. was supported in part by the NASA Hubble Fellowship grant HST-HF2-51467.001-A awarded by the Space Telescope Science Institute, which is operated by the Association of Universities for Research in Astronomy, Incorporated, under NASA contract NAS5-26555. E.H.P also received support from the Heising-Simons Foundation through the 51 Pegasi b Fellowship.

% References
\bibliography{report} % bibliography data in report.bib
\bibliographystyle{spiebib} % makes bibtex use spiebib.bst

\end{document}